# Dosimetric and Biologic Differences in Flattened and Flattening-Filter-Free Beam Treatment Plans

Yue Yan, PhD[*], Poonam Yadav, PhD[†,‡], Michael Bassetti, MD, PhD[†], Kaifang Du, PhD[†], Daniel Saenz, PhD[*,†], Paul Harari, MD[†] and Bhudatt R. Paliwal, PhD[*,†]

*Department of [*]Medical Physics and [†]Human Oncology, University of Wisconsin-Madison and [‡]University of Wisconsin, Riverview Cancer Center, Wisconsin Rapids, WI.*

**Purpose:** To quantitatively compare the dosimetric and biologic differences in treatment plans from flattened and flattening-filter-free (FFF) beam for three anatomic cancer sites.

**Methods and Materials:** Treatment plans with static intensity-modulated radiotherapy beams and volumetric modulated arc therapy beams were generated for 13 patients for both the flattened beam and the FFF beam of the TrueBeam system. Beam energies of 6 MV and 10 MV were chosen for planning. A total of 104 treatment plans were generated in 13 patients. In order to analyze the biological effectiveness of treatment plans, dose volume histograms (DVH) were utilized. Flattened and FFF beam plans are quantitatively compared.

**Results:** In head and neck cases, for VMAT plans, dose reduction in the FFF beam plans compared to the flattened beam in left cochlea, right submandibular gland and right parotid gland reached up to 2.36 Gy, 1.21 Gy and 1.45 Gy, respectively. Similarly, for static IMRT plans, the dose reduction of the FFF beam plans compared to the flattened beam plans for the same organs reached up to 0.34 Gy, 1.36 Gy and 1.46 Gy, respectively. Overall, for head and neck, the FFF beam plans achieved mean dose reduction of up to 5%, 7% and 9%, respectively for above organs at risk. For lung and prostate cases, the FFF beams provided lower or comparable NTCP values to organ-at-risk (OAR) compared to the flattened beam for all plans.

**Conclusions:** In general, we observed treatment plans utilizing FFF beams can improve dose sparing to OARs without compromising the target coverage. Significant dose sparing effect is obtained for head and neck cancer cases, especially for the cases with relatively large field sizes ($\approx 16 \times 20 \ cm^2$). For lung and prostate cases, compared to the flattened beam, the FFF beam based treatment plans provide lower or comparable dose to most OARs.

---

Acknowledgements-- Many thanks to Dr. Edward Bender, Dr. Bryan Bednarz in the University of Wisconsin Madison for their enthusiastic discussion on flattened beam and the FFF beam.
 Conflict of interest: none
Reprint requests to Bhudatt R. Paliwal, PhD, Departments of Human Oncology and Medical Physics, School of Medicine and Public Health, University of Wisconsin Madison, 600 Highland Ave, Madison, WI 53792. Tel: (608) 263-8514; E-mail: paliwal@humonc.wisc.edu

# Introduction

Intensity-modulated radiation therapy (IMRT) techniques have led to improved conformal dose delivery methods. Modern IMRT techniques include static step-and-shoot IMRT, rotational IMRT (e.g. volumetric modulated arc therapy (VMAT)(1) and helical Tomotherapy (2). In contrast to the 3D-conformal radiotherapy (3D-CRT), IMRT provides improved dose conformity to the target as well as steeper dose gradients to surrounding tissues, which may lead to better local tumor control (3). However, probability of increased dose to surrounding tissues is comparatively higher in IMRT than 3D-CRT. The typical ratio of monitor units (MU) between the IMRT plan and the 3D-CRT plan is in the range of 3 to 5. This contributes to higher leakage from the gantry head and consequently increased dose to normal tissues and whole body in general (4, 5). This undesirable dose is likely to result in higher normal tissue complication probability (NTCP) (6). Furthermore, IMRT tends to have a prolonged treatment time to deliver dose compared to 3D-CRT. Several studies have highlighted that these limitations of IMRT may lead to increased probability of radiation induced secondary cancer (4). It is therefore desirable to reduce the unnecessary scatter from the gantry head and shorten the treatment time for IMRT delivery. The removal of the flattening filter has been a logical choice to reduce the scatter.

The flattening filter was first introduced to provide flat dose profiles at a certain depth. The development of IMRT eliminates the need for a flattening filter in modern linear accelerator (linac) systems. In recent years, the application of the flattening-filter-free (FFF) photon beam has been studied extensively (5, 7-16). Forward peaked dose profile is the major characteristic of the FFF beam (17-22). Compared with the flattened beam, the FFF beam also has increased dose rate (8-12), reduced dose to OAR (12), reduced neutron contamination for high energy beams

(>15 MV) (23) and reduced uncertainty in dose calculation (8). Thus, clinical application of the FFF beam would lead to reduced treatment time (11) and secondary cancer risk induced by radiation (11, 13).

Several clinical comparative studies have investigated the differences between the flattened beam and the FFF beam (23-29). Most of these clinical comparison studies focused on the time efficiency obtained from the high dose rate of the FFF beam compared with the flattened beam. The dosimetric and biological differences between the flattened beam and the FFF beam for static IMRT and VMAT plans with typical ($\approx 10 \times 10\ cm^2$) and large field sizes ($\approx 16 \times 20\ cm^2$) are not well understood. In the presented study, 6 MV and 10 MV beams were selected to design the treatment plans. Three clinical sites were used to investigate the dosimetric and biologic differences treatment plans using the flattened beam and the FFF beam. Static IMRT and VMAT techniques were utilized in this study.

## Materials and Methods

*Patient selection*

Thirteen anonymized patients with three anatomical cancer sites: head and neck, lung and prostate were studied. A case number was assigned to refer to each anonymized patient. Standard clinical constrains were provided by the physician for planning target volume (PTV) and OARs. These were applied to generate the treatment plans. The dose prescriptions were selected from the typical dose range prescribed by the physicians. For head and neck cases, the dose prescription ranged from 60 Gy to 70 Gy at 2 - 2.5 Gy/fraction. For lung cases, the dose prescription ranged from 45 Gy to 60 Gy at 1.5 - 2 Gy/fraction. For prostate cases, dose prescriptions ranged from 70 Gy to 78 Gy at 2 - 2.5 Gy/fraction.

*RT planning techniques*

A Varian TrueBeam (Varian Medical Systems, Palo Alto, CA) linac was commissioned on the Eclipse$^{TM}$ treatment planning system (TPS) (version 10.0). An Anisotropic-Analytical-Algorithm (AAA) was used to calculate the dose for both static IMRT and VMAT plans. The dose grid in the calculation was 2.5 mm for all plans. Photon beam energies of 6 MV and 10 MV were selected for this study. Beam modalities included flattened and FFF beams. All treatment parameters such as iso-center position, beam angle, arc number and field size were set to be identical for the flattened and the FFF beam plans. High definition 120 leaf MLC (2.5 mm width in the center and 5 mm width in the peripheral) was used to generate all treatment plans.

For each patient, 8 treatment plans were generated, including 6 MV (10 MV) flattened static IMRT plan, 6 MV (10 MV) FFF static IMRT plan, 6 MV (10 MV) flattened VMAT plan and 6 MV (10 MV) FFF VMAT plan. In the FFF beam mode, the maximum dose rate increases from 600 MU/min to 1400 MU/min for the 6 MV and to 2400 MU/min for the 10 MV photon beam. For the VMAT plans with FFF beams, the TPS automatically selects the optimal dose rate during the optimization process. In our study, the optimal dose rates of the VMAT plans are lower than the maximum dose rates of the FFF beam for both the energies. For treatment plans with large field sizes (e.g. $\approx 16 \times 20 \ cm^2$ in case 2), the optimal dose rates of the VMAT plans were largely reduced (e.g. about 350 MU/min in case 2 for 6 MV beam) from the maximum dose rates of the FFF beam. For static IMRT plans, a constant dose rate of 600 MU/min was applied to design the treatment plans. This eliminated the influence of the speed limit of the multi-leaf collimator (MLC). For all treatment plans, the normal tissue fall-off function was set to be same for all plans. The optimization processes were repeated five times for all static IMRT and VMAT plans in order to get an optimal dose distribution.

DICOM files were exported from the Eclipse workstation for all cases. CERR (the Computational Environment for Radiotherapy Research(30)) was used to calculate dose-volume-histogram (DVH) from the DICOM file. In-house developed Matlab codes (Math Works, Natick, MA) were used to calculate the dosimetric and biological doses and to perform statistical analysis. The dosimetric results were benchmarked with the Eclipse software system for each case and acceptable agreement was observed.

*Treatment plan evaluation*

Target coverage and dose to OARs were analyzed to evaluate the treatment plans for all cases. For all treatment plans, 95% of the target volume was normalized to 95% of the dose prescription for evaluation and optimization. The criteria used to evaluate the target coverage included conformity index (CI), target coverage (TC), conformity number (CN) and gradient index (GI). These are defined as (31):

$$CI = TV_{95}/PTV_{95} \quad (1)$$

$$TC = PTV_{95}/PTV \quad (2)$$

$$CN = CI/TC \quad (3)$$

$$GI = TV_{50}/TV_{95} \quad (4)$$

In Eq. 1-Eq. 4, $TV_{95}$ and $PTV_{95}$ refer to the treated volume and the planning target volume (PTV) covered by the 95% dose line. A value closer to one indicates better target coverage for all indices. A paired sample t-test (**32**) was applied to analyze the statistical differences of target coverage among patients (statistical significance, $p \leq 0.05$).

Besides the physical dose, biological doses including biological effective dose (BED) (33) and equivalent uniform dose (EUD) (34-36)) were also calculated. Using Gay and Niemierko's model (36), tumor control probability (TCP) and normal tissue complication probability (NTCP) were calculated for targets and OARs respectively by using the calculated EUD. Biological parameters ($\alpha/\beta$, $TD_{50}$, $TCD_{50}$, $a$ and $\gamma_{50}$) were selected from the published references (36-41).

## Results

The dose distribution of one selected patient from each studied site is shown in Fig.1. The DVHs of one patient are selected from each treatment site and are shown in Fig 2 and Fig 3. Target coverage for the FFF beam is comparable to the flattened beam for both static IMRT and VMAT plans ($p > 0.05$). Significant differences between the flattened beam and the FFF beam were observed for the VMAT plans in lung cancer cases. For VMAT plans, the FFF beam provides a higher relative mean dose ($D_{mean}/D_x$) to the target compared with the flattened beam for both 6 MV and 10 MV beams. For the static IMRT plans, the difference between the flattened beam and the FFF beam is not significant. The target coverage analysis for head and neck and lung cases is shown in Table e1 and Table e2. For other treatment sites, the differences between the flattened beam and the FFF beam are not significant ($p \leq 0.05$).

Comparisons of dose to OARs between the flattened beam and the FFF beam are shown in Table 1- Table 3. Among the three cancer sites, the dose sparing effect of the FFF beam is significant in head and neck cases. For certain OARs such as left cochlea, larynx and right submandibular gland, noticeable dose sparing effect is obtained by the FFF beam compared to the flattened beam. In Table 1, for static IMRT plans, the FFF beam has the most significant dose sparing effect compared to the flattened beam on larynx and right submandibular gland. Compared to the flattened beam, the FFF beam reduces mean dose up to 2.05 Gy and 1.36 Gy for larynx and right submandibular, respectively, for 10 MV beam. For VMAT plans, left cochlea and larynx show the best dose sparing effect from the FFF beam compared to the flattened beam. The mean dose reductions come up to 2.36 and 2.82 Gy, respectively. Compared with the static IMRT plans, the VMAT plans show considerable differences between the flattened beam and the FFF beam. Relative dose ratio between the flattened beam and the FFF beam for five head and neck cases are shown in Table 3e. For left cochlea, larynx and right submandibular gland, the mean dose reduction of the FFF beam compared to the flattened beam reaches up to 5%, 3% and 5%, respectively.

For the lung cancer case, as shown in Fig 2- Fig 3, larynx has the most significant dose sparing effect from the FFF beam compared to the flattened beam, both for 10 MV static IMRT and VMAT plans. In Table 2, the reduction in the mean dose for the FFF beam compared to the flattened beam is 1.6 Gy for larynx. For other organs, comparable doses are obtained by the FFF beam and flattened beam for static IMRT and VMAT plans in both beam energies. Relative dose ratio between the flattened beam and the FFF beam for four lung cases is shown in Table 4e. For organs such as heart and lungs, the FFF beam compared to the flattened beam, tends to provide higher maximum dose of 2% and 3%, respectively. , This effect is more significant in the 10 MV VMAT lung plans compared with the three other plans.

For the prostate cancer, for both VMAT and static IMRT plans, the FFF beam provides a comparable or improved dose sparing effect to OARs. The dosimetric and biological dose and NTCP values of a selected prostate case are shown in Table 3. The maximum reduction in the mean dose is obtained for the right hip (1.03 Gy) in the 10 MV VMAT plan compared to the flattened beam. The relative dose ratio between the flattened beam and the FFF beam for 4 patients is shown in Table 5e. For rectum, in VMAT plans, the FFF beam provided slightly higher (1%) maximum dose compared to the flattened beam.. For all OAR, the FFF beam provided improved dose sparing effect and NTCP values compared to the flattened beam.

## Discussion

Overall, in our clinical comparison of the head and neck cases the differences between the FFF beam and the flattened beam are significant for both 6 MV and 10 MV beams. For lung and prostate, results were comparable. Head and neck cases required relatively larger field size to cover the target. As shown in Fig 1, large field sizes ($\sim 16 \times 20 \ cm^2$) were used to cover the target. For the VMAT plans, two arcs with different iso-centers were used to provide the required dose coverage for the PTVs. In other cancer sites, typical field sizes ($\approx 10 \times 10 \ cm^2$) were used to create the treatment plans.

The noticeable dose sparing effect of the FFF beam compared with the flattened beam for large treatment field size is due to the forward peak beam profiles of the FFF beam. There is no observable difference between the beam profiles of flattened beam and FFF beam for small field size (e.g. $6 \times 6 \ cm^2$). For relatively large field sizes (e.g. $16 \times 20 \ cm^2$), the FFF beam provided lower dose to the out-of-field region compared with the flattened beam for both 6 MV and 10 MV beams. This is of clinical significance for cases receiving a high radiation dose (~ 70 Gy) and having a diversity of sensitive normal tissue structures as found in the head and neck region. When we increased the beam energy from 6 MV to 10 MV, the dose reduction effect in the out-of-field region was significant for the FFF beam compared to the flattened beam. This fact also explains the improved dose sparing effect of the FFF beam in head and neck cases in both 10 MV static IMRT and VMAT plans compared to the 6 MV plans.

However, due to the non-uniform beam profile of the FFF beam, compared to the flattened beam, the FFF beam tends to use more MUs to deliver the uniform dose to target. In our clinical investigation, the ratio of MUs between the FFF beam plans and the flattened beam plans was typically around 1.3 for the static IMRT and VMAT plans. The higher MUs of the FFF beam may lead to increased dose leakage from the MLC and escalated dose to OARs. Based on our investigation, even with the increased leakage dose from MLC, the FFF beam can still provide comparable dose sparing effect to OARs in most cases.

Randomized studies have documented dosimetric improvement to OARs using IMRT has dramatically improved overall toxicity outcomes and quality of life for patients (42-44). Despite these improvements, both acute and late toxicity represent on going challenges to successful head and neck cancer treatments. In the cases we examined, the lower dose to submandibular glands and parotid glands could all contribute to lower xerostomia rates as the mean dose to each of these OARs has been directly associated with xerostomia (45). In the dose range where xerostomia (45) is likely, a linear correlation with the mean dose is apparent suggesting even modest dose improvement may have a clinical impact. Other toxicities that may be affected include voice quality, swallowing function, breathing function and cataract development. In head and neck, for both static IMRT and VMAT plans, the flattened beam and the FFF beam provided improved dose sparing to right parotid gland and right submandibular gland. Clearly, the FFF beam provided a significant dose sparing effect to the right submandibular gland compared to the flattened beam and should lead to a lower probability of xerostomia. The higher larynx dose associated with a flattened beam may contribute to poor voice quality, as the larynx mean dose has been correlated with laryngeal edema (47). The late toxicity of radiation treatment is directly related to both the overall dose and the dose per fraction. The lower mean and maximal doses achieved to OAR reduces both the total delivered dose to several critical organs, as well as the daily fraction size. Thus, FFF treatment for the examined head and neck cases may allow for the same TCP coverage with a decreased risk of late side effects of treatment.

## Conclusion

In this manuscript, 13 clinical cancer cases for three anatomical sites were investigated to analyze the dosimetric and biologic differences between the flattened and the FFF beam plans in terms of target coverage and dose to OARs. It was observed that the FFF beam provides comparable target coverage to the flattened beam in all 3 sites of cancer. The FFF beam for head and neck cancer obtained observable dose sparing. For other two sites, the FFF beam provided improved dose sparing effect to most of the OARs. For certain OARs such as the heart in the VMAT plan for lung cancer, the FFF beam delivered higher maximum dose. Due to the speed limit of the MLC, the maximum dose rate of the FFF beam may be considerably lower than the theoretical maximum dose rate value, especially for large field sizes.

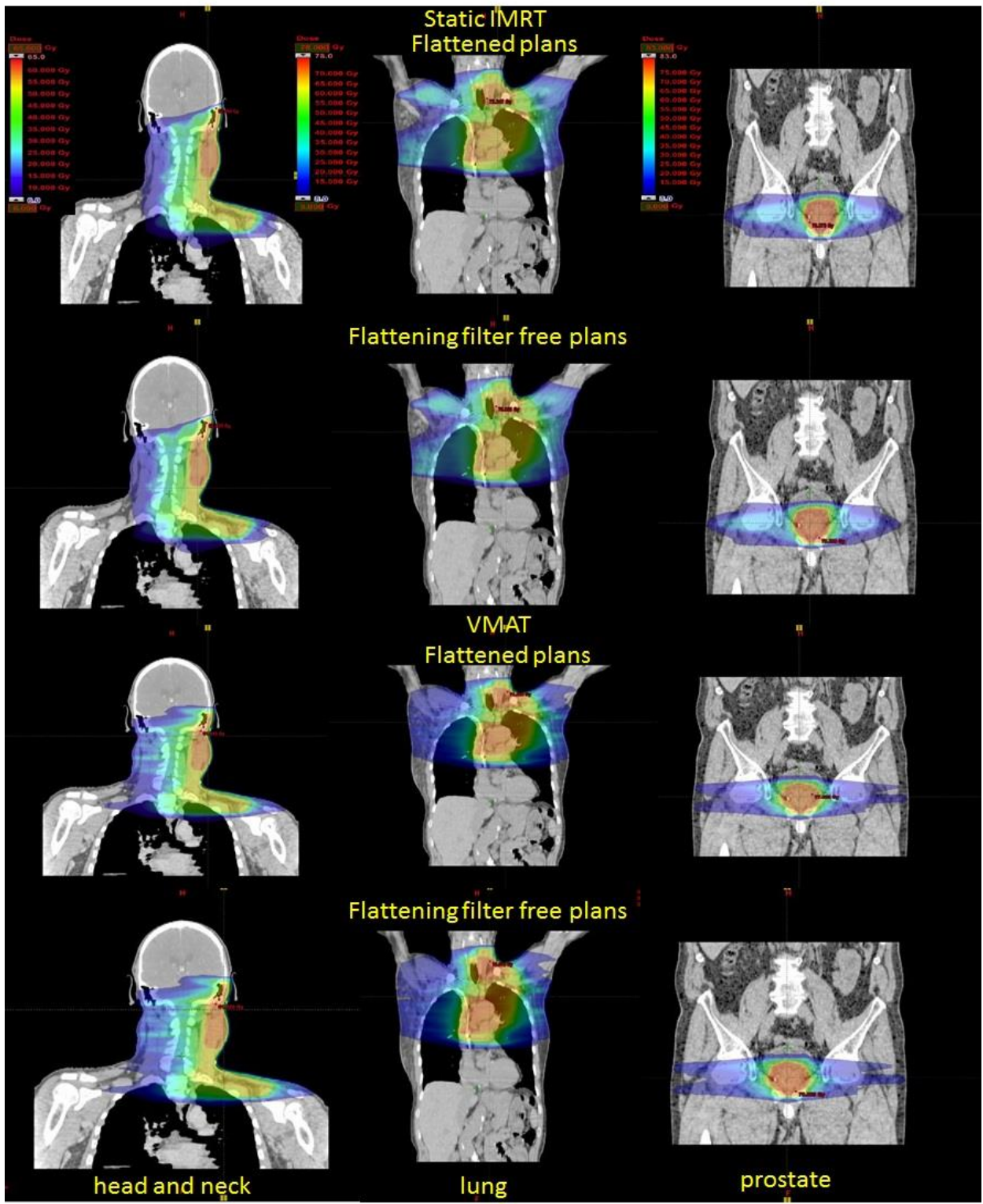

**Fig. 1.** Dose distribution for 6 MV head and neck (left column), lung (middle column) and prostate (right column) patients for one plan per case. From top to bottom, it shows the static IMRT flattened plans, static IMRT FFF plans, VMAT flattened plans and VMAT FFF plans.

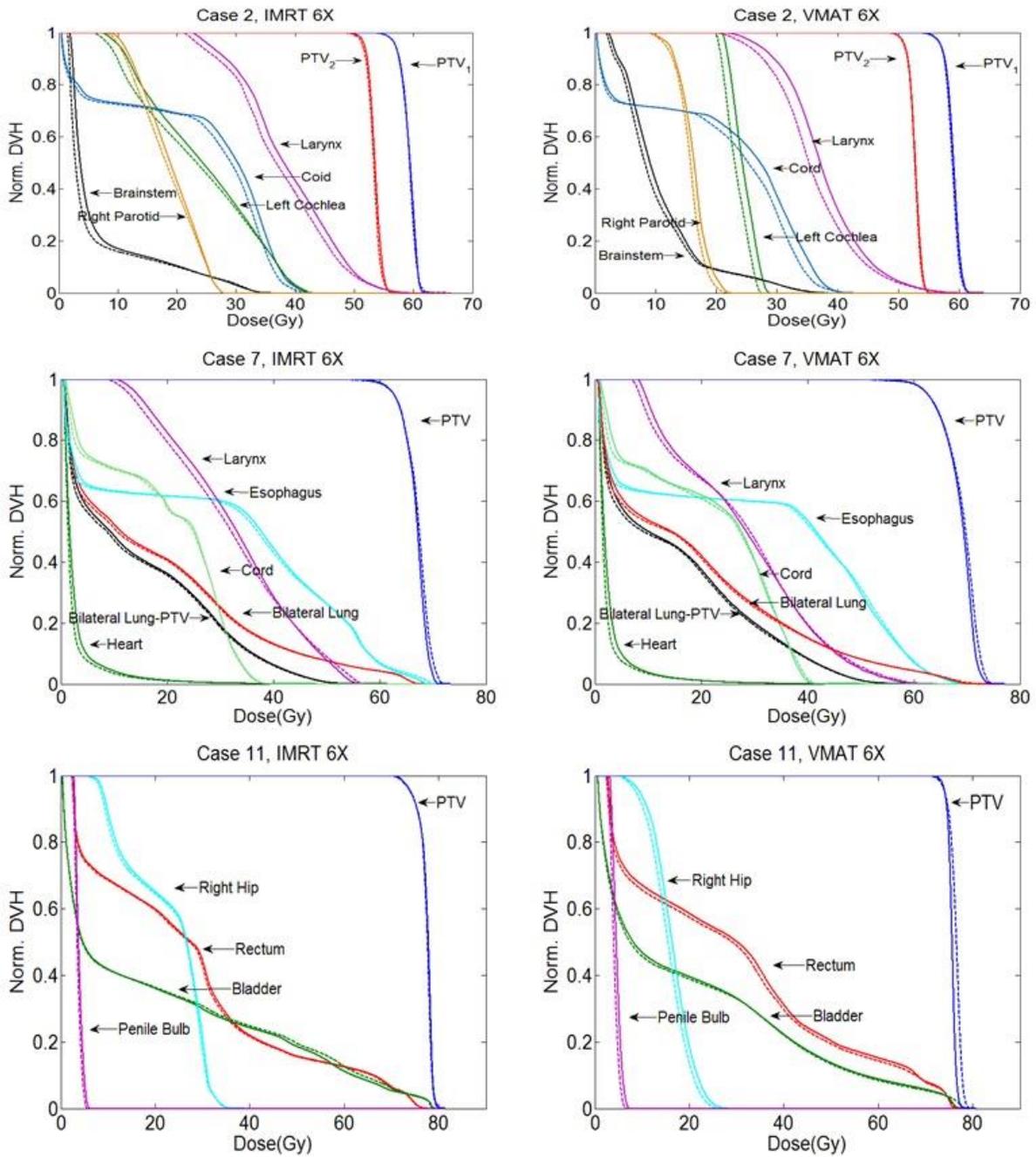

**Fig 2.** Normalized treatment plans comparison between the flattened and the FFF beams for the static IMRT and the VMAT plans for beam energy 6 MV. Selected cases include head and neck case, lung case and prostate case. The solid lines are the flattened beam plans and the dashed lines are the FFF beam plans.

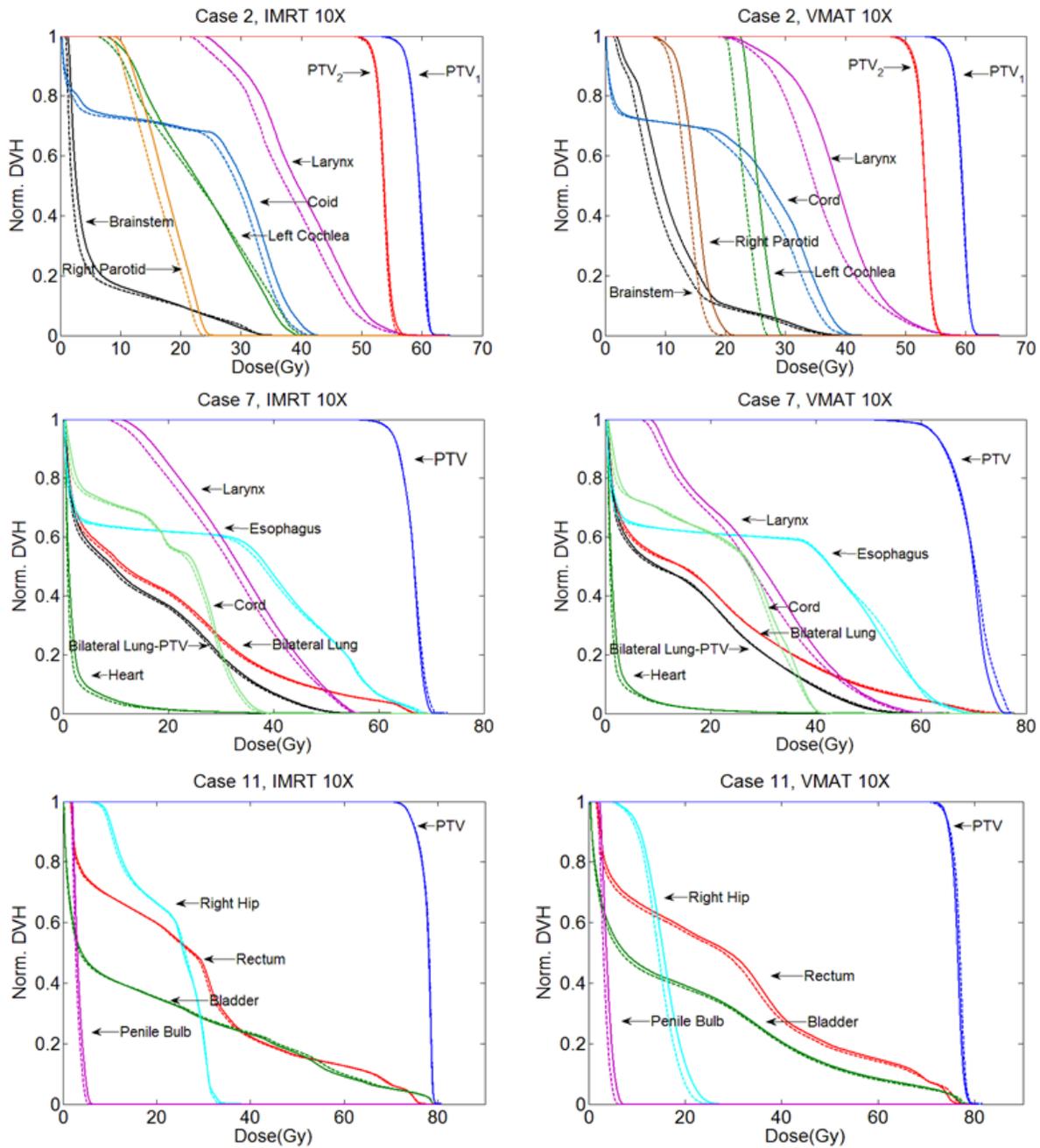

**Fig. 3.** Normalized treatment plans comparison between the flattened and the FFF beams for the static IMRT and the VMAT plans for beam energy 10 MV. Selected cases include head and neck case and lung case and prostate case. The solid lines are the flattened beam plans and the dashed lines are the FFF beam plans.

**Table 1** Physical dose, biological dose and NTCP values for OARs in the head and neck case #2.

| OAR | Flattened Beam | | | | FFF Beam | | | |
|---|---|---|---|---|---|---|---|---|
| | Max Dose (Gy) | Mean Dose (Gy) | Mean BED (Gy) | NTCP | Max Dose (Gy) | Mean Dose (Gy) | Mean BED (Gy) | NTCP |
| **Dose Prescription/Fraction Number: 60 Gy/30 fx** | | | | | | | | |
| **6 MV IMRT** | | | | | | | | |
| Left Cochlea | 42.88 | 24.28 | 31.85 | 1.02E-04 | 42.92 | 23.32 | 30.52 | 9.42E-05 |
| Larynx | 59.18 | 38.72 | 54.02 | 2.53E-04 | 59.38 | 37.49 | 51.91 | 2.03E-04 |
| Right Parotid | 27.81 | 18.80 | 22.98 | 2.94E-09 | 27.99 | 18.21 | 22.18 | 1.56E-09 |
| Right Submandibular | 29.95 | 23.92 | 30.33 | 1.77E-07 | 28.71 | 23.41 | 29.55 | 1.19E-07 |
| Brainstem | 35.94 | 6.92 | 8.12 | 8.45E-08 | 35.94 | 6.18 | 7.30 | 9.11E-08 |
| Cord | 42.50 | 23.84 | 32.35 | 1.96E-06 | 42.10 | 22.90 | 30.84 | 1.12E-06 |
| **10 MV IMRT** | | | | | | | | |
| Left Cochlea | 39.45 | 23.37 | 30.22 | 2.33E-05 | 40.67 | 23.03 | 29.90 | 4.00E-05 |
| Larynx | 59.21 | 39.84 | 55.95 | 3.03E-04 | 58.75 | 37.78 | 52.37 | 1.65E-04 |
| Right Parotid | 25.38 | 17.39 | 20.93 | 7.11E-10 | 24.18 | 15.94 | 18.94 | 1.35E-10 |
| Right Submandibular | 28.68 | 23.36 | 29.47 | 1.16E-07 | 26.97 | 22.01 | 27.43 | 3.60E-08 |
| Brainstem | 35.15 | 6.12 | 7.20 | 6.14E-08 | 35.35 | 5.44 | 6.48 | 7.95E-08 |
| Cord | 42.79 | 24.09 | 32.82 | 2.61E-06 | 42.02 | 23.00 | 31.05 | 1.19E-06 |
| **6 MV VMAT** | | | | | | | | |
| Left Cochlea | 28.85 | 24.48 | 31.17 | 2.72E-07 | 27.49 | 23.28 | 29.35 | 1.09E-07 |
| Larynx | 59.80 | 37.65 | 51.97 | 1.27E-04 | 59.41 | 36.04 | 49.25 | 9.64E-05 |
| Right Parotid | 22.37 | 16.23 | 19.22 | 2.18E-10 | 21.96 | 15.64 | 18.42 | 1.11E-10 |
| Right Submandibular | 26.75 | 22.83 | 28.68 | 7.42E-08 | 26.29 | 22.40 | 28.03 | 5.06E-08 |
| Brainstem | 37.47 | 10.88 | 12.71 | 1.29E-07 | 38.38 | 10.23 | 11.93 | 1.43E-07 |
| Cord | 42.57 | 21.59 | 28.82 | 8.00E-07 | 42.03 | 20.46 | 26.99 | 4.40E-07 |
| **10 MV VMAT** | | | | | | | | |
| Left Cochlea | 29.74 | 25.62 | 32.95 | 5.60E-07 | 27.21 | 23.26 | 29.30 | 9.40E-08 |
| Larynx | 59.16 | 38.57 | 53.63 | 1.56E-04 | 58.91 | 35.76 | 48.81 | 1.01E-04 |
| Right Parotid | 21.56 | 15.08 | 17.66 | 5.56E-11 | 19.68 | 13.63 | 15.74 | 8.82E-12 |
| Right Submandibular | 26.77 | 22.33 | 27.93 | 4.91E-08 | 26.46 | 21.12 | 26.13 | 1.82E-08 |
| Brainstem | 39.07 | 11.44 | 13.50 | 2.96E-07 | 39.11 | 10.01 | 11.71 | 1.90E-07 |
| Cord | 42.80 | 21.99 | 29.59 | 1.27E-06 | 41.56 | 20.58 | 27.27 | 5.25E-07 |

**Table 2** Physical dose, biological dose and NTCP values for OARs in the lung case #7.

| OAR | Flattened Beam | | | | FFF Beam | | | |
|---|---|---|---|---|---|---|---|---|
| | Max Dose (Gy) | Mean Dose (Gy) | Mean BED (Gy) | NTCP | Max Dose (Gy) | Mean Dose (Gy) | Mean BED (Gy) | NTCP |
| **6 MV IMRT** | | | | | | | | |
| Cord | 39.69 | 19.99 | 25.48 | 1.27E-07 | 39.60 | 19.69 | 25.10 | 1.16E-07 |
| Esophagus | 69.57 | 30.37 | 45.20 | 5.77E-02 | 70.73 | 30.11 | 44.89 | 8.36E-02 |
| Heart | 48.18 | 2.76 | 2.99 | 6.74E-12 | 47.81 | 2.37 | 2.55 | 3.79E-12 |
| Larynx | 56.84 | 32.83 | 43.61 | 1.38E-05 | 56.91 | 32.09 | 42.59 | 1.61E-05 |
| Lungs | 69.30 | 18.26 | 24.97 | 6.51E-05 | 70.15 | 17.91 | 24.52 | 5.62E-05 |
| **10 MV IMRT** | | | | | | | | |
| Cord | 39.79 | 19.86 | 25.34 | 1.38E-07 | 38.70 | 19.30 | 24.54 | 8.90E-08 |
| Esophagus | 68.57 | 30.48 | 45.49 | 4.65E-02 | 69.21 | 30.14 | 45.00 | 5.64E-02 |
| Heart | 47.69 | 2.36 | 2.58 | 9.26E-12 | 47.84 | 1.97 | 2.15 | 5.43E-12 |
| Larynx | 56.57 | 33.50 | 44.64 | 1.67E-05 | 56.44 | 31.90 | 42.22 | 1.12E-05 |
| Lungs | 68.80 | 18.37 | 25.22 | 7.04E-05 | 69.35 | 17.93 | 24.61 | 5.78E-05 |
| **6 MV VMAT** | | | | | | | | |
| Cord | 43.33 | 21.93 | 28.67 | 9.19E-07 | 43.83 | 21.66 | 28.38 | 1.11E-06 |
| Esophagus | 70.93 | 31.49 | 47.29 | 4.76E-02 | 71.89 | 31.38 | 47.23 | 5.31E-02 |
| Heart | 49.66 | 2.82 | 3.04 | 3.29E-12 | 50.69 | 2.44 | 2.64 | 4.34E-12 |
| Larynx | 63.50 | 28.08 | 36.60 | 5.38E-06 | 64.35 | 28.01 | 36.64 | 7.01E-06 |
| Lungs | 71.97 | 18.97 | 26.12 | 9.32E-05 | 74.08 | 18.62 | 25.69 | 8.15E-05 |
| **10 MV VMAT** | | | | | | | | |
| Cord | 43.20 | 21.88 | 28.71 | 1.08E-06 | 42.23 | 21.35 | 27.91 | 8.92E-07 |
| Esophagus | 71.43 | 31.89 | 48.27 | 5.66E-02 | 70.64 | 32.06 | 48.71 | 4.79E-02 |
| Heart | 49.38 | 2.34 | 2.54 | 3.95E-12 | 48.88 | 2.03 | 2.22 | 5.46E-12 |
| Larynx | 62.93 | 29.30 | 38.41 | 7.67E-06 | 62.83 | 27.70 | 36.01 | 4.69E-06 |
| Lungs | 74.65 | 19.01 | 26.33 | 9.94E-05 | 77.76 | 19.05 | 26.52 | 1.05E-04 |

Dose Prescription/Fraction Number: 66 Gy/33 fx

**Table 3** Physical dose, biological dose and NTCP values for OARs in the prostate case #11.

| | Dose Prescription/Fraction Number: 78 Gy/39 fx | | | | | | | |
|---|---|---|---|---|---|---|---|---|
| | Flattened Beam | | | | FFF Beam | | | |
| OAR | Max Dose (Gy) | Mean Dose (Gy) | Mean BED (Gy) | NTCP | Max Dose (Gy) | Mean Dose (Gy) | Mean BED (Gy) | NTCP |
| | 6 MV IMRT | | | | | | | |
| Rectum | 77.68 | 27.46 | 37.94 | 1.82E-03 | 77.71 | 27.14 | 37.46 | 1.70E-03 |
| Bladder | 79.39 | 20.41 | 29.24 | 1.89E-04 | 79.39 | 20.79 | 29.98 | 2.79E-04 |
| Right Hip | 38.93 | 22.75 | 27.78 | 4.66E-07 | 38.89 | 22.40 | 27.31 | 4.02E-07 |
| Left Hip | 35.64 | 24.25 | 29.79 | 8.31E-07 | 36.26 | 23.88 | 29.30 | 7.84E-07 |
| Penile Bulb | 6.06 | 3.92 | 4.06 | 3.13E-21 | 5.58 | 3.64 | 3.76 | 8.68E-22 |
| | 10 MV IMRT | | | | | | | |
| Rectum | 77.43 | 27.46 | 38.04 | 1.86E-03 | 77.40 | 27.18 | 37.62 | 1.85E-03 |
| Bladder | 79.14 | 19.17 | 27.26 | 8.51E-05 | 79.52 | 19.39 | 27.73 | 1.13E-04 |
| Right Hip | 38.09 | 22.97 | 28.02 | 4.16E-07 | 36.79 | 22.83 | 27.84 | 3.93E-07 |
| Left Hip | 34.76 | 24.49 | 30.04 | 6.81E-07 | 35.01 | 24.28 | 29.80 | 7.27E-07 |
| Penile Bulb | 6.27 | 3.36 | 3.46 | 8.53E-22 | 5.56 | 2.97 | 3.06 | 1.14E-22 |
| | 6 MV VMAT | | | | | | | |
| Rectum | 77.23 | 29.86 | 42.17 | 2.82E-03 | 78.22 | 29.00 | 40.87 | 2.73E-03 |
| Bladder | 79.19 | 21.04 | 29.31 | 8.11E-05 | 80.80 | 20.69 | 28.84 | 7.25E-05 |
| Right Hip | 28.26 | 16.41 | 18.85 | 3.40E-10 | 27.05 | 15.53 | 17.73 | 1.43E-10 |
| Left Hip | 29.86 | 15.69 | 17.93 | 2.60E-10 | 29.53 | 14.82 | 16.84 | 1.32E-10 |
| Penile Bulb | 7.25 | 4.68 | 4.87 | 5.85E-20 | 6.53 | 4.15 | 4.31 | 8.55E-21 |
| | 10 MV VMAT | | | | | | | |
| Rectum | 77.76 | 29.15 | 41.12 | 2.51E-03 | 78.14 | 28.31 | 39.88 | 2.88E-03 |
| Bladder | 78.80 | 20.23 | 28.15 | 5.60E-05 | 79.79 | 19.68 | 27.37 | 4.38E-05 |
| Right Hip | 27.14 | 15.61 | 17.82 | 1.48E-10 | 25.28 | 14.58 | 16.51 | 3.81E-11 |
| Left Hip | 29.34 | 14.72 | 16.71 | 9.85E-11 | 29.27 | 14.46 | 16.39 | 8.74E-11 |
| Penile Bulb | 7.14 | 3.94 | 4.08 | 9.35E-21 | 6.06 | 3.30 | 3.40 | 5.34E-22 |

**Table e1.** Target coverage analysis for head and neck cases (n=5).

| | Head and Neck | | | | | |
|---|---|---|---|---|---|---|
| | VMAT | | | | | |
| | 6 MV | | | 10 MV | | |
| Parameter | Flattened | FFF | p | Flattened | FFF | p |
| $D_{mean}/D_x$ | 0.99±0.03 | 0.99±0.02 | 0.9 | 1.00±0.02 | 1.00±0.02 | 0.65 |
| CI | 1.20±0.15 | 1.28±0.29 | 0.31 | 1.42±0.36 | 1.46±0.45 | 0.62 |
| CN | 1.26±0.16 | 1.34±0.30 | 0.31 | 1.50±0.38 | 1.54±0.47 | 0.62 |
| GI | 10.79±10.60 | 8.83±7.08 | 0.29 | 8.40±7.20 | 7.44±5.47 | 0.29 |
| TCP | 0.84±0.14 | 0.84±0.14 | 0.75 | 0.85±0.14 | 0.85±0.14 | 0.35 |
| | IMRT | | | | | |
| Parameter | Flattened | FFF | p | Flattened | FFF | p |
| $D_{mean}/D_x$ | 0.99±0.01 | 0.99±0.02 | 0.08 | 0.99±0.01 | 0.99±0.01 | 0.32 |
| CI | 1.26±0.14 | 1.27±0.15 | 0.21 | 1.22±0.11 | 1.22±0.11 | 0.82 |
| CN | 1.32±0.15 | 1.34±0.16 | 0.21 | 1.29±0.12 | 1.29±0.12 | 0.82 |
| GI | 10.80±10.98 | 10.42±10.53 | 0.13 | 11.03±11.67 | 10.43±10.60 | 0.28 |
| TCP | 0.84±0.14 | 0.84±0.14 | 0.13 | 0.84±0.15 | 0.84±0.15 | 0.44 |

**Table e2.** Target coverage analysis for lung cases (n=4).

| | Lung | | | | | |
|---|---|---|---|---|---|---|
| | VMAT | | | | | |
| | 6 MV | | | 10 MV | | |
| Parameter | Flattened | FFF | p | Flattened | FFF | p |
| $D_{mean}/D_x$ | 1.01±0.02 | 1.02±0.02 | 0.01 | 1.03±0.02 | 1.04±0.01 | 0.04 |
| CI | 1.11±0.11 | 1.12±0.11 | 0.06 | 1.12±0.12 | 1.13±0.10 | 0.91 |
| CN | 1.17±0.12 | 1.18±0.12 | 0.06 | 1.18±0.12 | 1.19±0.10 | 0.92 |
| GI | 3.30±0.18 | 3.27±0.16 | 0.1 | 3.14±0.19 | 3.12±0.19 | 0.29 |
| TCP | 0.71±0.06 | 0.72±0.06 | 0.34 | 0.72±0.06 | 0.72±0.06 | 0.03 |
| | IMRT | | | | | |
| Parameter | Flattened | FFF | p | Flattened | FFF | p |
| $D_{mean}/D_x$ | 1.01±0.01 | 1.01±0.01 | 0.86 | 1.01±0.02 | 1.01±0.02 | 0.77 |
| CI | 1.15±0.03 | 1.16±0.03 | 0.12 | 1.14±0.03 | 1.14±0.02 | 0.89 |
| CN | 1.21±0.03 | 1.22±0.03 | 0.12 | 1.20±0.03 | 1.20±0.03 | 0.89 |
| GI | 3.38±0.40 | 3.37±0.41 | 0.46 | 3.25±0.42 | 3.19±0.40 | 0.07 |
| TCP | 0.71±0.06 | 0.71±0.06 | 0.59 | 0.71±0.06 | 0.71±0.06 | 0.89 |

**Table e3.** Relative dose and NTCP ratio (FFF/flattened) to OARs for head and neck cases (n=5).

| | VMAT | | | | | |
|---|---|---|---|---|---|---|
| | **6 MV** | | | **10 MV** | | |
| **OAR** | **Mean Dose Ratio** | **Max Dose Ratio** | **NTCP Ratio** | **Mean Dose Ratio** | **Max Dose Ratio** | **NTCP Ratio** |
| Left Cochlea | 0.97±0.02 | 0.96±0.02 | 0.56±0.22 | 0.95±0.06 | 0.94±0.04 | 0.53±0.51 |
| Larynx | 0.98±0.02 | 1.00±0.01 | 0.87±0.23 | 0.97±0.04 | 1.00±0.02 | 0.89±0.25 |
| Cord | 0.96±0.01 | 1.00±0.02 | 0.73±0.25 | 0.98±0.03 | 1.00±0.03 | 0.97±0.60 |
| Brainstem | 0.87±0.11 | 0.90±0.13 | 0.50±0.47 | 0.86±0.09 | 0.91±0.11 | 0.45±0.37 |
| Right parotid | 0.91±0.06 | 0.95±0.04 | 0.25±0.22 | 0.92±0.03 | 0.95±0.03 | 0.21±0.14 |
| Right submandibular | 0.94±0.04 | 0.98±0.01 | 0.53±0.13 | 0.93±0.02 | 0.98±0.04 | 0.48±0.34 |

| | IMRT | | | | | |
|---|---|---|---|---|---|---|
| | **6 MV** | | | **10 MV** | | |
| **OARs** | **Mean Dose Ratio** | **Max Dose Ratio** | **NTCP Ratio** | **Mean dose Ratio** | **Max dose Ratio** | **NTCP Ratio** |
| Left cochlea | 0.97±0.01 | 0.98±0.02 | 0.74±0.25 | 0.96±0.04 | 0.99±0.06 | 0.98±1.04 |
| Larynx | 0.98±0.01 | 1.00±0.00 | 0.98±0.12 | 0.98±0.02 | 1.01±0.02 | 0.94±0.29 |
| Cord | 0.97±0.01 | 0.98±0.01 | 0.70±0.16 | 0.97±0.02 | 0.99±0.01 | 0.71±0.19 |
| Brainstem | 0.91±0.02 | 0.96±0.06 | 0.73±0.40 | 0.88±0.05 | 0.92±0.09 | 0.56±0.48 |
| Right parotid | 0.97±0.02 | 1.01±0.01 | 0.53±0.29 | 0.95±0.04 | 0.98±0.02 | 0.45±0.46 |
| Right submandibular | 0.97±0.02 | 0.98±0.02 | 0.74±0.17 | 0.97±0.03 | 0.98±0.03 | 0.72±0.37 |

**Table e4.** Relative dose and NTCP ratio (FFF/flattened) to OARs for lung cases (n=4).

| | VMAT | | | | | |
|---|---|---|---|---|---|---|
| | 6 MV | | | 10 MV | | |
| OARs | Mean dose ratio | Max dose ratio | NTCP ratio | Mean dose ratio | Max dose ratio | NTCP ratio |
| Cord | 0.99±0.01 | 1.01±0.01 | 1.20±0.18 | 0.98±0.01 | 1.00±0.02 | 0.92±0.2 |
| Esophagus | 0.99±0.01 | 1.01±0.00 | 1.10±0.05 | 1.00±0.01 | 1.01±0.02 | 1.18±0.31 |
| Heart | 0.94±0.06 | 1.01±0.02 | 1.08±0.25 | 0.95±0.06 | 1.02±0.02 | 1.25±0.16 |
| Larynx | 0.95±0.06 | 1.00±0.02 | 0.84±0.65 | 0.94±0.00 | 1.01±0.02 | 0.76±0.21 |
| Lungs | 0.99±0.01 | 1.01±0.02 | 0.94±0.06 | 1.00±0.01 | 1.03±0.02 | 1.00±0.17 |
| | IMRT | | | | | |
| | 6 MV | | | 10 MV | | |
| OARs | Mean dose ratio | Max dose ratio | NTCP ratio | Mean dose ratio | Max dose ratio | NTCP ratio |
| Cord | 0.98±0.02 | 1.00±0.01 | 0.85±0.21 | 0.97±0.01 | 0.98±0.01 | 0.65±0.17 |
| Esophagus | 0.99±0.01 | 1.01±0.01 | 1.14±0.21 | 0.98±0.00 | 1.01±0.00 | 1.06±0.12 |
| Heart | 0.94±0.06 | 1.00±0.01 | 0.84±0.19 | 0.92±0.06 | 1.01±0.00 | 0.76±0.16 |
| Larynx | 0.94±0.05 | 0.99±0.02 | 0.79±0.53 | 0.92±0.04 | 0.98±0.02 | 0.48±0.26 |
| Lungs | 0.98±0.01 | 1.01±0.01 | 0.88±0.06 | 0.98±0.00 | 1.01±0.00 | 0.83±0.02 |

**Table e5.** Relative dose and NTCP ratio (FFF/flattened) to OARs for prostate cases (n=4).

| | VMAT | | | | | |
| --- | --- | --- | --- | --- | --- | --- |
| | 6 MV | | | 10 MV | | |
| OARs | Mean dose ratio | Max dose ratio | NTCP ratio | Mean dose ratio | Max dose ratio | NTCP ratio |
| Rectum | 0.99±0.01 | 1.01±0.01 | 1.10±0.11 | 0.99±0.02 | 1.00±0.00 | 1.07±0.06 |
| Bladder | 0.96±0.04 | 1.02±0.01 | 0.71±0.34 | 0.98±0.02 | 1.01±0.00 | 0.91±0.23 |
| Right Hip | 0.97±0.02 | 0.97±0.02 | 0.57±0.23 | 0.96±0.04 | 0.96±0.04 | 0.52±0.42 |
| Left Hip | 0.97±0.02 | 0.99±0.01 | 0.64±0.24 | 0.97±0.04 | 0.99±0.03 | 0.79±0.48 |
| Penile Bulb | 0.86±0.07 | 0.87±0.08 | 0.12±0.08 | 0.87±0.02 | 0.89±0.03 | 0.10±0.03 |
| | IMRT | | | | | |
| | 6 MV | | | 10 MV | | |
| OARs | Mean dose ratio | Max dose ratio | NTCP ratio | Mean dose ratio | Max dose ratio | NTCP ratio |
| Rectum | 0.98±0.01 | 1.00±0.00 | 0.95±0.02 | 0.99±0.00 | 1.00±0.00 | 0.98±0.03 |
| Bladder | 0.99±0.02 | 1.00±0.00 | 0.98±0.38 | 0.99±0.02 | 1.00±0.00 | 0.94±0.29 |
| Right Hip | 0.98±0.02 | 1.00±0.01 | 0.84±0.29 | 0.98±0.01 | 0.98±0.02 | 0.84±0.22 |
| Left Hip | 0.98±0.02 | 1.00±0.02 | 0.86±0.24 | 0.99±0.01 | 1.00±0.02 | 0.87±0.21 |
| Penile Bulb | 0.92±0.02 | 0.92±0.03 | 0.25±0.07 | 0.89±0.02 | 0.9±0.04 | 0.19±0.15 |

## References

1. Yu CX. Intensity-modulated arc therapy with dynamic multileaf collimation: an alternative to tomotherapy.*Phys Med Biol*1995;40:1435-1449.
2. Mackie TR, Holmes T, Swerdloff S,et al. Tomotherapy: a new concept for the delivery of dynamic conformal radiotherapy. *Med Phys*1993; 20:1709–1719.
3. Cahlon O, Hunt M, Zelefsky MJ, Intensity-modulated radiation therapy: supportive data for prostate cancer.Semin Radiat Oncol 2008;18:48–57.
4. Followill D, Geis P, Boyer A. Estimates of whole-body dose equivalent produced by beam intensity modulated conformal therapy, *Int J Radiat Oncol Biol Phys*1997;38:667–672.
5. Cashmore J, Ramtohul M, Ford D. Lowering whole-body radiation doses in pediatric intensity- modulated radiotherapy through the use of unflattened photon beams.*Int J Radiat Oncol Biol Phys*2011;80:1220–1227.
6. Diallo I, Haddy N, Adjadj E, et al. Frequency distribution of second solid cancer locations in relation to the irradiated volume among 115 patients treated for childhood cancer.*Int J Radiat Oncol Biol Phys* 2009;74:876–883.
7. Ong CL, Dahele M, Slotman BJ, et al. Dosimetric Impact of the Interplay Effect During Stereotactic Lung Radiation Therapy Delivery Using Flattening Filter-Free Beams and Volumetric Modulated Arc Therapy.*Int J Radiat Oncol Biol Phys*2013;86:743–748.
8. Georg D, Knöös T, McClean B. Current status and future perspective of flattening filter free photon beams. *Med Phys*2011;38:1280–1293.
9. Huang Y, Siochi RA, Bayouth JE. Dosimetric properties of a beam quality-matched 6 MV unflattened photon beam. *J Appl Clin Med Phys*2012;13:71-81.
10. Vassiliev ON, Titt U, Kry SF, et al. Monte Carlo study of photon fields from a flattening filter-free clinical accelerator.*Med Phys* 2006;33:820-827.
11. Reggiori G, Mancosu P, Castiglioni S, et al. Can volumetric modulated arc therapy with flattening filter free beams play a role in stereotactic body radiotherapy for liver lesions? A volume-based analysis. *Med Phys* 2012;39:1112–8.
12. Titt U, Vassiliev ON, Ponisch F, et al. A flattening filter free photon treatment concept evaluation with Monte Carlo. *Med Phys* 2006;33:1595.
13. Kry SF, Vassiliev ON, Mohan R. Out-of-field photon dose following removal of the flattening filter from a medical accelerator. *Phys Med Biol*2010;55:2155–66.
14. Kragl G, Baier F, Lutz S, et al. Flattening filter free beams in SBRT and IMRT: dosimetric assessment of peripheral doses.*Z Med Phys* 2011;21:91–101.
15. Tsiamas P, Seco J, Han Z, et al. A modification of flattening filter free linac for IMRT. *Med Phys* 2011;38:2342-52.
16. Wang Y, Khan MK, Ting JY, el al. Surface dose investigation of the flattening filter- free photon beams.*Int J Radiat Oncol Biol Phys* 2012;83: e281–5.
17. Vassiliev ON, Titt U, Pönisch F, et al. Dosimetric properties of photon beams from a flattening filter free clinical accelerator.*Phys Med Biol*2006;51:1907–17.
18. Ponisch F, Titt U, Vassiliev ON, et al. Properties of unflattened photon beams shaped by a multileaf collimator. *Med Phys*2006;33:1738-46.
19. Kragl G, Albrich D,Georg D. Radiation therapy with unflattened photon beams: dosimetric accuracy of advanced dose calculation algorithms. *Radiother Oncol*2011;100:417–23.
20. Stevens SW, Rosser KE, Bedford JL. A 4 MV flattening filter-free beam: commissioning and application to conformal therapy and volumetric modulated arc therapy.*Phys Med Biol*2011;56:3809–24.
21. Chang Z, Wu Q, Adamson J, et al. Commissioning and dosimetric characteristics of TrueBeam system: composite data of three TrueBeam machines. *Med Phys*2012;39 :6981–7018.
22. Hrbacek J, Lang S, Klöck S. Commissioning of photon beams of a flattening filter-free

linear accelerator and the accuracy of beam modeling using an anisotropic analytical algorithm. *Int J Radiat Oncol Biol Phys* 2011;80:1228–37.
23. Kry SF, Titt U, Pönisch F, et al. Reduced neutron production through use of a flattening-filter-free accelerator.*Int J Radiat Oncol Biol Phys* 2007;68:1260–4.
24. Thomas EM, Popple RA, Prendergast BM, et al. Effects of flattening filter-free and volumetric-modulated arc therapy delivery on treatment efficiency.*J Appl Clin Med Phys* 2013;14:4328.
25. Zhang GG, Ku L, Dilling TJ, et al. Volumetric modulated arc planning for lung stereotactic body radiotherapy using conventional and unflattened photon beams: a dosimetric comparison with 3D technique.*Radiat Oncol*2011;6:152.
26. Nicolini G, Ghosh-Laskar S, Shrivastava SK, et al. Volumetric modulation arc radiotherapy with flattening filter-free beams compared with static gantry IMRT and 3D conformal radiotherapy for advanced esophageal cancer: a feasibility study.*Int J Radiat Oncol Biol Phys* 2012;84:553–60.
27. Ong CL, Dahele M, Cuijpers JP, et al. Dosimetric impact of intrafraction motion during RapidArc stereotactic vertebral radiation therapy using flattened and flattening filter-free beams.*Int J Radiat Oncol Biol Phys*2013;86;420-5.
28. Prendergast BM, Fiveash JB, Popple RA, et al. Flattening filter-free linac improves treatment delivery efficiency in stereotactic body radiation therapy.*J Appl Clin Med Phys*2013;14: 4126.
29. Vassiliev ON, Kry SF, Kuban DA, et al. Treatment-planning study of prostate cancer intensity-modulated radiotherapy with a Varian Clinac operated without a flattening filter.*Int J Radiat Oncol Biol Phys*2007;68:1567–71.
30. Deasy JO, Blanco AI, and Clark VH.CERR: A computational environment for radiotherapy research.*Med Phys*2003;30: 979-85.
31. Dhabaan A, Elder E, Schreibmann E, et al.Dosimetric performance of the new high-definition multileaf collimator for intracranial stereotactic radiosurgery. *J Appl Clin Med Phys*2010;11:3040
32. Student. The probable error of a mean. *Biometrika*1908;6:1-25.
33. Fowler JF.The linear-quadratic formula and progress in fractionated radiotherapy.*BJR*1989;62: 679-94.
34. Niemierko A.Reporting and analyzing dose distributions: A concept of equivalent uniform dose. *Med Phys*1997;24:103-10.
35. Niemierko A.A generalized concept of equivalent uniform dose (EUD).*Med Phys*1999;26:1100.
36. Gay HA and Niemierko A.A free program for calculating EUD-based NTCP and TCP in external beam radiotherapy.*Phys Medcia* 2007;23:115-125.
37. Oinam AS, Singh L, Shukla A. Dose volume histogram analysis and comparison of different radiobiological models using in-house developed software. *J Med Phys* 2011;36:220-9.
38. Okunie P, Morgan D, Niemierko A, et al.Radiation dose-response of human tumors. *Int J Radiat Oncol Biol Phys*1995;32:1227-37.
39. Emami B, Lyman J, BrownA, et al.Tolerance of normal tissue to therapeutic irradiation. *Int J Radiat Oncol Biol Phys*1991;21:109-122.
40. Mukesh MB, Harris E, Collette S, et al.Normal tissue complication probability (NTCP) parameters for breast fibrosis: Pooled results from two randomised trials.*Radiother Oncol*2013;108: 293-298.
41. Svolou P, Tsougos I, Theodorou K,et al.The Use of radiobiological parameters and the evaluation of NTCP models. How do they affect the ability to estimate radiationinduced complications? *WC 2009, IFMBE Proceedings* 25/3, 2009, pp 292-294.
42. Kam MK, Leung SF, Zee B, *et al*. Prospective randomized study of intensity-modulated radiotherapy on salivary gland function in early-stage nasopharyngeal carcinoma patients.*J*

*Clin Oncol* 2007;25:4873-9.
43. Pow EH, Kwong DL, McMillan AS, *et al*. Xerostomia and quality of life after intensity-modulated radiotherapy vs. conventional radiotherapy for early-stage nasopharyngeal carcinoma: initial report on a randomized controlled clinical trial. *Int J Radiat Oncol Biol Phys* 2006;66:981-91.
44. Nutting C M, Morden J P, Harrington K J, *et al*. Parotid-sparing intensity modulated versus conventional radiotherapy in head and neck cancer (PARSPORT): a phase 3 multicentre randomised controlled trial. *Lancet Oncol* 2011;12:127-36.
45. Little M, Schipper M, Feng FY, *et al*. Reducing xerostomia after chemo-IMRT for head-and-neck cancer: beyond sparing the parotid glands. *Int J Radiat Oncol Biol Phys* 2011;83:1007–1014.
46. Anand A K, Jain J, Negi P S, *et al*. Can dose reduction to one parotid gland prevent xerostomia?--A feasibility study for locally advanced head and neck cancer patients treated with intensity-modulated radiotherapy. *Clin Oncol* 2006; 18(6), 497-504.
47. Sanguineti G, Adapala P, Endres EJ, *et al*. Dosimetric predictors of laryngeal edema. *Int J Radiat Oncol Biol Phys* 2007;68:741-9.